\documentclass[aps,pra,amsmath,two column,amssymb,showpacs]{revtex4}
\usepackage{txfonts}
\usepackage{mathbbold}
\usepackage{epsfig,bm,dcolumn}
\usepackage{graphicx}
\usepackage{color}
\usepackage{overpic}

\begin{document}

\title{BCS-BEC quantum phase transition and collective excitations  in two-dimensional Fermi gases\\ with $p$- and $d$-wave pairings}

\author{Gaoqing Cao,$^1$ Lianyi He,$^2$\footnote{Email address: lianyi@th.physik.uni-frankfurt.de} and Pengfei Zhuang$^1$}

\address{1 Department of Physics, Tsinghua University, Beijing 100084, China\\
2 Frankfurt Institute for Advanced Studies and Institute for Theoretical Physics, J. W. Goethe
University, 60438 Frankfurt am Main,
Germany}

\date{\today}

\begin{abstract}
It is generally believed that the BCS-BEC evolution in fermionic systems with $s$-wave pairing is a smooth crossover. However, for nonzero orbital-angular-momentum pairing such as $p$- or $d$-wave pairing, the system undergoes a quantum phase transition at the point
where the chemical potential $\mu$ vanishes. In this paper, we study the BCS-BEC quantum phase transition and the collective excitations associated with the order-parameter fluctuations in two-dimensional fermionic systems with $p$- and $d$-wave pairings. We show that the quantum phase transition in such systems can be generically traced back to the infrared behavior of the fermionic excitation at $\mu=0$: $E_{\bf k}\sim k^l$, where $l=1,2$ is the quantum number of the orbital angular momentum. The nonanalyticity of the thermodynamic quantities is due to the infrared divergence caused by the fermionic excitation at $\mu=0$. As a result, the evolution of the Anderson-Bogoliubov mode is not smooth: Its velocity is nonanalytical across the quantum phase transition.

\end{abstract}


\maketitle

\section{Introduction}

It was proposed by Eagles~\cite{Eagles} and Leggett~\cite{Leggett} several decades ago that, by tuning the strength of the attractive interaction in a many-fermion system, one can realize an evolution from the Bardeen-Cooper-Schrieffer (BCS) superfluidity to Bose-Einstein condensation (BEC) of difermion molecules. For $s$-wave interaction, it is generally believed that the BCS-BEC evolution is a smooth crossover~\cite{BCSBEC01,BCSBEC02,BCSBEC03,BCSBEC04,BCSBEC05,BCSBEC06,BCSBEC07,BCSBEC08}.  One of the most interesting systems is the dilute spin-$\frac{1}{2}$ Fermi gas in three dimensions with short-range $s$-wave attraction. The system is characterized by a dimensionless parameter $1/(k_{\rm F}a_s)$, where $a_s$ is the $s$-wave scattering length of the short-range interaction and $k_{\rm F}$ is the Fermi momentum in the absence of interaction. For such a system, one finds a smooth BCS-BEC crossover when the parameter $1/(k_{\rm F}a_s)$ goes from $-\infty$ to $\infty$. In addition, the Anderson-Bogoliubov collective mode of fermionic superfluidity with weak attraction evolves smoothly to the Bogoliubov excitation of weakly repulsive Bose condensate with strong attraction~\cite{BCSBEC03,collective,lattice02}. The smooth BCS-BEC crossover with $s$-wave interaction has been experimentally studied by using ultracold fermionic atoms~\cite{BCSBECexp}, where the $s$-wave scattering length was tuned by means of the Feshbach resonance.

On the other hand, for nonzero orbital-angular-momentum pairing, such as $p$- or $d$-wave pairing~\cite{PIP,QPT01,QPT02,QPT03,QPT04,QPT05}, it is generally accepted that the BCS-BEC evolution is not smooth but associated with some quantum phase transition. Such quantum phase transition cannot be characterized by a change of symmetry or the associated order parameter. Instead, different quantum phases can be distinguished topologically~\cite{PIP}. The quantum phase transitions and the finite temperature phase diagrams of single-species polarized Fermi gases tuned across a $p$-wave Feshbach resonance were studied by Gurarie, Radzihovsky, and Andreev \cite{QPT04} (for an extensive study, see Ref. \cite{BCSBEC08}). In three dimensions, they showed that across the $p$-wave Feshbach resonance, the system would undergo a quantum phase transition from a $p_x$-wave to a $p_x+ip_y$-wave superfluid. Furthermore, the latter state undergoes a topological transition at zero chemical potential $\mu$.

In this paper, we explore some generic aspects of the BCS-BEC quantum phase transition in two-dimensional (2D) $p$- and $d$-wave paired Fermi superfluids, which is expected to occur at the point where the chemical potential $\mu$ vanishes~\cite{PIP,QPT01,QPT02}. While for nonzero orbital-angular-momentum pairing it is important to explore the order parameter symmetry, in this paper we show that the generic nature of the BCS-BEC quantum phase transition at $\mu=0$ in two-dimensional fermionic systems is independent of the order parameter symmetry but depends on the infrared behavior of the order parameter.

In general, the superfluid order parameter or the gap function $\Delta({\bf k})$ is momentum dependent for nonzero orbital-angular-momentum pairing. The crucial observation in this paper is that the infrared behavior of the order parameter $\Delta({\bf k})$ depends only on the quantum number $l$ of the orbital angular momentum. For $l$th-wave pairing, we have
\begin{eqnarray}
\Delta({\bf k})\sim k^l,\ \ \ \ \ {\bf k}\rightarrow 0.
\end{eqnarray}
For instance, the order parameter of $p$-wave pairing can be expressed as $\Delta({\bf k})=\Delta_0 kg(\varphi)$ where $\varphi$ is the polar angle in two-dimensions. The anisotropic $p_x$ ($p_y$) pairing corresponds to $g(\varphi)=\cos\varphi$ ($\sin\varphi$) and the complex $p_x+ip_y$ pairing corresponds to $g(\varphi)=e^{i\varphi}$. For any case, we find that the infrared behavior of the order parameter is $\Delta({\bf k})\sim k$, independent of the order parameter symmetry.

The single-particle excitation spectrum in the superfluid state reads $E_{\bf k}=[\xi_{\bf k}^2+|\Delta({\bf k})|^2]^{1/2}$, where $\xi_{\bf k}={\bf k}^2/(2M)-\mu$, with $M$ being the fermion mass. Therefore, for $p$-wave ($l=1$) and $d$-wave ($l=2$) pairings, the infrared behavior of the single-particle spectrum at the quantum phase transition point $\mu=0$ is solely determined by the order parameter, that is
\begin{eqnarray}
E_{\bf k}(\mu=0)\sim k^l,\ \ \ \ \ {\bf k}\rightarrow 0.
\end{eqnarray}
Note that the above behavior applies only to $l=1$ and $l=2$. For $l>2$, the infrared behavior becomes dominated by the kinetic term $\xi_{\bf k}$. In this paper, we show that the type of 2D momentum integral ($m,n$ integer)
\begin{eqnarray}\label{IRINT}
{\cal I}(\mu)=\int_0^{\infty}k dk \frac{k^n}{E_{\bf k}^m}
\end{eqnarray}
appears in the expressions of various physical quantities. This integral is divergent in infrared at $\mu=0$ if $lm\geq n+2$. Therefore, for a 2D system with nonzero orbital-angular-momentum pairing, nonanalyticities generally appear due to the infrared divergence at the quantum critical point $\mu=0$.

The nonanalyticities of the thermodynamic quantities at vanishing chemical potential which signal the quantum phase transitions in 2D $p$- and $d$-wave paired superfluids were first studied by Botelho and Sa de Melo using Nozieres--Schmitt-Rink (NSR) type potentials \cite{QPT01,QPT02}. In this paper, we extend their conclusions significantly. We show that the nonanalyticities are essentially related to the infrared behavior of the interaction potential. Therefore, the quantum phase transition at vanishing chemical potential is independent of the details of the interaction potential as well as the order parameter symmetry, which can be intuitively understood by the integral (\ref{IRINT}). We also show that the evolution of the Anderson-Bogoliubov mode is also not smooth; that is, the sound velocity goes nonanalytically across the quantum phase transition.

In the rest of this paper, we study the BCS-BEC quantum phase transition and the behavior of Anderson-Bogoliubov mode across the phase transition for two typical cases: $p$-wave pairing in spinless Fermi gases in Sec. II and $d$-wave pairing in spin-$\frac{1}{2}$ Fermi gases in Sec. III.

\section{$p$-wave pairing in spinless Fermi gases}
For a 2D interaction potential $V(r)$, the momentum-space matrix element $V({\bf k},{\bf k}^\prime)$ can be expressed as \cite{QPT01}
\begin{eqnarray}
V({\bf k},{\bf k}^\prime)=\sum_{n=-\infty}^{\infty} e^{in\theta_{{\bf k}{\bf k}^\prime}}V^{(n)}(k,k^\prime),
\end{eqnarray}
where $\theta_{{\bf k}{\bf k}^\prime}$ is the angle between ${\bf k}$ and ${\bf k}^\prime$ and the coefficients $V^{(n)}(k,k^\prime)$ are given by
\begin{eqnarray}
V^{(n)}(k,k^\prime)=2\pi\int_0^\infty dr r J_n(kr) J_n(k^\prime r)V(r).
\end{eqnarray}
Here $J_n(x)$ is the Bessel function of order $n$. The $n=\pm l$ components correspond to the $l$th angular momentum channel.

It is possible to retain only the $n=\pm l$ terms, by assuming that the dominant contribution to the scattering processes between fermions occurs
in the $l$th angular momentum channel. This assumption may be experimentally relevant since atom-atom dipole interactions split different angular momentum channels such that they may be tuned independently.

Using the properties of the Bessel function, in the low energy limit $k\rightarrow 0$ and $k^\prime \rightarrow 0$ we have
\begin{eqnarray}
V^{(\pm l)}(k,k^\prime)\sim k^l (k^\prime)^l.
\end{eqnarray}
Therefore, the potential becomes separable in the low energy limit. While it is certainly not separable for general values of $k$ and $k^\prime$,
in the following we use separable potential for the $l$th wave interaction to simplify our formulation. Since the generic features of the BCS-BEC quantum phase transition with nonzero orbital-angular-momentum pairing are related solely to the infrared behavior of the interaction potential, the use of a separable potential is without loss of generality.

The many-body Hamiltonian of 2D spinless fermions can be written as $H=H_0+H_{\rm int}$, where the single-particle part reads
$H_0=\sum_{\bf k}\xi_{\bf k}\psi^{\dagger}_{\bf k}\psi^{\phantom{\dag}}_{\bf k}$ and the $p$-wave interaction part can be written as
\begin{eqnarray}
H_{\rm int}=\sum_{{\bf k},{\bf k}^\prime,{\bf q}}V_p({\bf k},{\bf k}^\prime)\psi^\dagger_{{\bf k}+{\bf q}/2}\psi^\dagger_{-{\bf k}+{\bf q}/2}\psi^{\phantom{\dag}}_{-{\bf k}^\prime+{\bf q}/2}\psi^{\phantom{\dag}}_{{\bf k}^\prime+{\bf q}/2}.
\end{eqnarray}
The generic infrared behavior of the $p$-wave interaction potential is $V_p({\bf k},{\bf k}^\prime)\sim kk^\prime$ for $k,k^\prime\rightarrow 0$.  Without loss of generality, we consider a separable potential for the $p$-wave interaction,  $V_p({\bf k},{\bf k}^\prime)=
-\lambda_p\Gamma_p({\bf k})\Gamma_p^*({\bf k}^\prime)$, where $\lambda_p$ is the coupling constant and the function $\Gamma_p(\bf k)$ characterizes the $p$-wave pairing symmetry. The infrared behavior of the function $\Gamma_p(\bf k)$ is
\begin{eqnarray}
\Gamma_p({\bf k})\sim k,\ \ \ \ \ {\bf k}\rightarrow 0.
\end{eqnarray}

In the functional path integral formalism, the partition function of the system at finite temperature $T$ is given by ${\cal Z}=\int \mathcal{ D} \psi \mathcal{D}\psi^\dagger\exp\left\{-{\cal S}[\psi,\psi^\dagger]\right\}$, where the action reads
\begin{eqnarray}
{\cal S}[\psi,\psi^\dagger]=\int_0^{1/T} d\tau\left[\sum_{\bf k}\psi_{\bf k}^\dagger(\tau)\partial_\tau \psi^{\phantom{\dag}}_{\bf k}(\tau)+ H(\psi,\psi^\dagger)\right].
\end{eqnarray}
To decouple the interaction term we introduce an auxiliary complex pairing field $\Phi_{\bf q}(\tau)$, which couples to $\psi^\dagger\psi^\dagger$, and apply the Hubbard-Stratonovich transformation. Using the Nambu-Gor$'$kov basis
$\Psi_{\bf k}=(\psi^{\phantom{\dag}}_{\bf k},\psi^\dagger_{\bf k})^{\rm T}$ and integrating out the fermionic degrees of freedom, we obtain
$\mathcal {Z}=\int\mathcal{D} \Phi \mathcal{D} \Phi^{\ast}\exp\big\{-{\cal S}_{\rm{eff}}[\Phi, \Phi^{\ast}]\big\}$, with the effective action given by
\begin{equation}
{\cal S}_{\rm{eff}} = \int_0^{1/T}d\tau \left[\sum_{\bf q}\frac{|\Phi_{\bf q}(\tau)|^{2}}{\lambda_p}
 +\frac{1}{2}\sum_{\bf k,k^\prime}\left(\xi_{\bf k}\delta_{\bf k,k^\prime}- \mbox{Trln}{\bf G}_{\bf k,k^\prime}^{-1}\right)\right],
\label{effaction}
\end{equation}
where the inverse single-particle Green's function ${\bf G}_{\bf k,k^\prime}^{-1}$ is given by
\begin{eqnarray}
{\bf G}_{\bf k,k^\prime}^{-1}=\left(\begin{array}{cc}-(\partial_{\tau}+\xi_{\bf k})\delta_{\bf k,k^\prime} & \Phi_{{\bf k}-{\bf k}^\prime}(\tau)\Gamma_p(\frac{{\bf k}+{\bf k}^\prime}{2})\\
\Phi^*_{{\bf k}-{\bf k}^\prime}(\tau)\Gamma_p^*(\frac{{\bf k}+{\bf k}^\prime}{2})& -(\partial_{\tau}-\xi_{\bf k})\delta_{\bf k,k^\prime}\end{array}\right).
\end{eqnarray}

The effective action ${\cal S}_{\rm{eff}}[\Phi, \Phi^*]$ cannot be evaluated precisely. In this work, we consider mainly the zero-temperature
case. Therefore, we follow the conventional approach to the BCS-BEC crossover problem; that is, we first consider the superfluid ground state which
corresponds to the saddle point of the effective action and then study the Gaussian fluctuation around the saddle point. The Gaussian-fluctuation
part corresponds to the collective modes, including the gapless Goldstone mode or the Anderson-Bogoliubov mode. This Goldstone mode appears as a resonance in the spectrum of the density-density correlation function \cite{Griffin}. Therefore, it can be experimentally probed by many techniques, including two-photon Bragg scattering \cite{Bragg,lattice01}.

In the superfluid ground state, the pairing field $\Phi$ acquires a nonzero expectation value $\Delta_0$, which can be set to be real without loss of generality. Then we decompose the pairing field as $\Phi_{\bf q}(\tau)=\Delta_0\delta_{\bf q,0}+\phi_{\bf q}(\tau)$, where $\phi$ is the fluctuation around the mean field. The effective action ${\cal S}_{\rm{eff}}[\Phi,\Phi^*]$ can be expanded in powers of the fluctuation $\phi$, that is,
\begin{eqnarray}
{\cal S}_{\rm{eff}}[\Phi,\Phi^*]={\cal S}_{\rm{eff}}^{(0)}(\Delta_0)+{\cal S}_{\rm{eff}}^{(2)}[\phi,\phi^*]+\cdots,
\label{sexpansion}
\end{eqnarray}
where ${\cal S}_{\rm{eff}}^{(0)}(\Delta_0)$ is the saddle point or mean-field effective action with $\Delta_0$
determined by the saddle point condition $\partial{\cal S}_{\rm{eff}}^{(0)}/\partial\Delta_0=0$.

\subsection{Quantum phase transition}

Neglecting the beyond-mean-field contribution, which is generally thought to be small, we obtain the effective potential
\begin{eqnarray}
{\cal V}(\Delta_0,\mu)=\frac{\Delta_0^2}{\lambda_p}+\frac{1}{2}\sum_{\bf k}
\left(\xi_{\bf k}-E_{\bf k}\right)
\end{eqnarray}
at zero temperature, where the single-particle excitation spectrum $E_{\bf k}$ is given by
\begin{eqnarray}
E_{\bf k}=\sqrt{\xi_{\bf k}^2+|\Delta({\bf k})|^2}.
\end{eqnarray}
Here $\Delta({\bf k})=\Delta_0\Gamma_p({\bf k})$ is the gap function for $p$-wave pairing. The order parameter $\Delta_0$ is determined by
minimizing the effective potential, which gives the gap equation
\begin{eqnarray}
\frac{1}{\lambda_p}=\sum_{\bf k}\frac{|\Gamma_p({\bf k})|^2}{4E_{\bf k}}.
\end{eqnarray}

To study the nature of the quantum phase transition, we focus on the thermodynamic potential $\Omega(\mu)\equiv {\cal V}(\Delta_0(\mu),\mu)$.
Note that the thermodynamic variable here is the chemical potential $\mu$. The order parameter $\Delta_0$ should be solved as an implicit
function of $\mu$ through the gap equation. The first derivative of the thermodynamic potential with respect to $\mu$ gives the number equation,
\begin{eqnarray}
n(\mu,\Delta_0)=-\frac{\partial \Omega(\mu)}{\partial\mu}=\frac{1}{2}\sum_{\bf k}
\left(1-\frac{\xi_{\bf k}}{E_{\bf k}}\right).
\end{eqnarray}
Then we consider the second derivative
\begin{eqnarray}
\alpha(\mu)=-\frac{\partial^2\Omega(\mu)}{\partial\mu^2}.
\end{eqnarray}
Considering the fact that $\Delta_0$ is an implicit function of $\mu$, we obtain
\begin{eqnarray}
\alpha(\mu)=\frac{\partial n(\mu,\Delta_0)}{\partial\mu}+\frac{\partial n(\mu,\Delta_0)}{\partial\Delta_0}\frac{\partial \Delta_0}{\partial\mu}.
\end{eqnarray}
The derivative of $\Delta_0$ with respect to $\mu$ can be obtained from the gap equation. We have
\begin{eqnarray}
\frac{\partial \Delta_0}{\partial\mu}=\frac{\partial n(\mu,\Delta_0)}{\partial\Delta_0}\left(\frac{\partial^2{\cal V}(\Delta_0,\mu)}{\partial\Delta_0^2}\right)^{-1}.
\end{eqnarray}
Therefore, $\alpha(\mu)$ can be expressed as
\begin{eqnarray}
\alpha(\mu)=\frac{\partial n(\mu,\Delta_0)}{\partial\mu}+\left(\frac{\partial n(\mu,\Delta_0)}{\partial\Delta_0}\right)^2
\left(\frac{\partial^2{\cal V}(\Delta_0,\mu)}{\partial\Delta_0^2}\right)^{-1}.
\end{eqnarray}
We note that the second term was missing in the previous study~\cite{QPT01}. However, the nonanalytical behavior is dominated by the first term. Therefore, the conclusions in Ref. \cite{QPT01} are still reliable. The derivatives in Eq. (20) can be explicitly
evaluated as
\begin{eqnarray}
&&\frac{\partial n(\mu,\Delta_0)}{\partial\mu}=\frac{\Delta_0^2}{2}\sum_{\bf k}\frac{|\Gamma_p({\bf k})|^2}{E_{\bf k}^3},\nonumber\\
&&\frac{\partial n(\mu,\Delta_0)}{\partial\Delta_0}=\frac{\Delta_0}{2}\sum_{\bf k}\frac{\xi_{\bf k}|\Gamma_p({\bf k})|^2}{E_{\bf k}^3},\nonumber\\
&&\frac{\partial^2{\cal V}(\Delta_0,\mu)}{\partial\Delta_0^2}=\frac{\Delta_0^2}{2}\sum_{\bf k}\frac{|\Gamma_p({\bf k})|^4}{E_{\bf k}^3}.
\end{eqnarray}

By using an NSR-type potential, Botelho and Sa de Melo found that the quantity $\alpha$ (proportional to the isothermal compressibility) is nonanalytical at $\mu=0$ \cite{QPT01}. Here we show that this nonanalyticity generally appears due to infrared divergence at $\mu=0$. Actually, for ${\bf k}\rightarrow 0$, we have $\Gamma_p({\bf k})\sim k$ and $E_{\bf k}\sim k$ at $\mu=0$. Therefore the momentum integrals in Eq. (18) are infrared safe at $\mu=0$. Then we further consider the derivatives of $\alpha(\mu)$ with respect to $\mu$. Actually, without explicit calculations, we
find that the following momentum integral
\begin{eqnarray}
{\cal I}_1(\mu)=\sum_{\bf k}\frac{|\Gamma_p({\bf k})|^2}{E_{\bf k}^5}
\end{eqnarray}
appears in the expression of $\partial^2\alpha(\mu)/\partial\mu^2$. At vanishing chemical potential $\mu=0$, the infrared behavior of the integral is
\begin{eqnarray}
{\cal I}_1(0)\sim\int_0^\epsilon kdk\frac{k^2}{k^5}\sim \int_0^\epsilon \frac{dk}{k^2}.
\end{eqnarray}
Therefore, this integral is infrared divergent. As a result, the derivative $\partial^2\alpha(\mu)/\partial\mu^2$ is divergent at the point $\mu=0$, which indicates that $\partial\alpha(\mu)/\partial\mu$ is discontinuous at $\mu=0$. Thus the quantity $\alpha(\mu)$ itself is nonanalytical across the quantum critical point $\mu=0$.

The above discussion shows that a quantum phase transition occurs at $\mu=0$, indicated by the nonanalyticity of the thermodynamic potential. The nonanalyticity is caused solely by the infrared behavior of the $p$-wave pairing potential and is independent of the details of the interaction as well as the order parameter symmetry (i.e., the angle dependence of the gap function). Therefore, the BCS-BEC quantum phase transition is quite generic in two-dimensional systems, driven by the infrared divergence at $\mu=0$.

\begin{figure}
\centering
\begin{overpic}
[scale=.92]{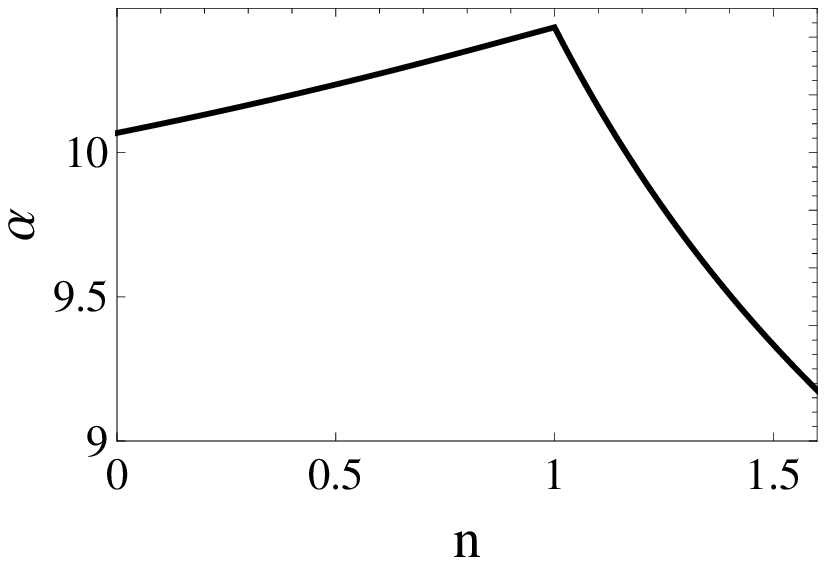}
\put(18,17){\includegraphics[scale=.35]{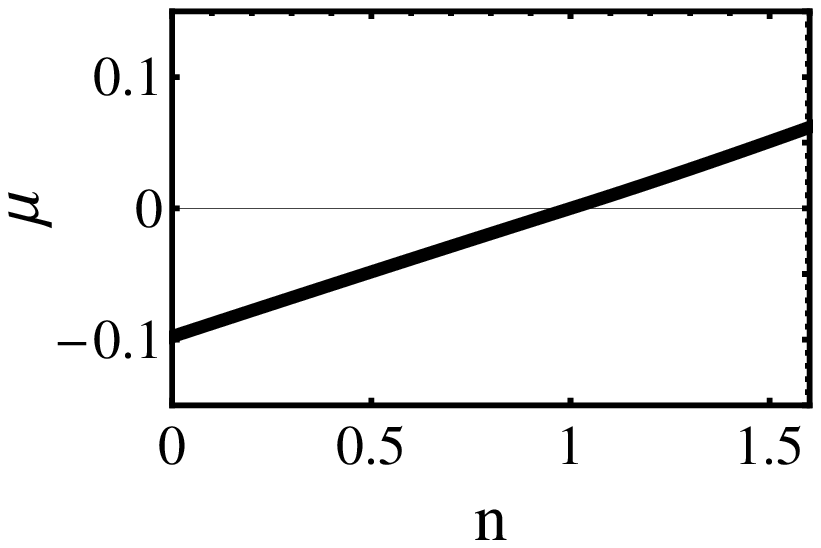}}
\end{overpic}
\begin{overpic}
[scale=.92]{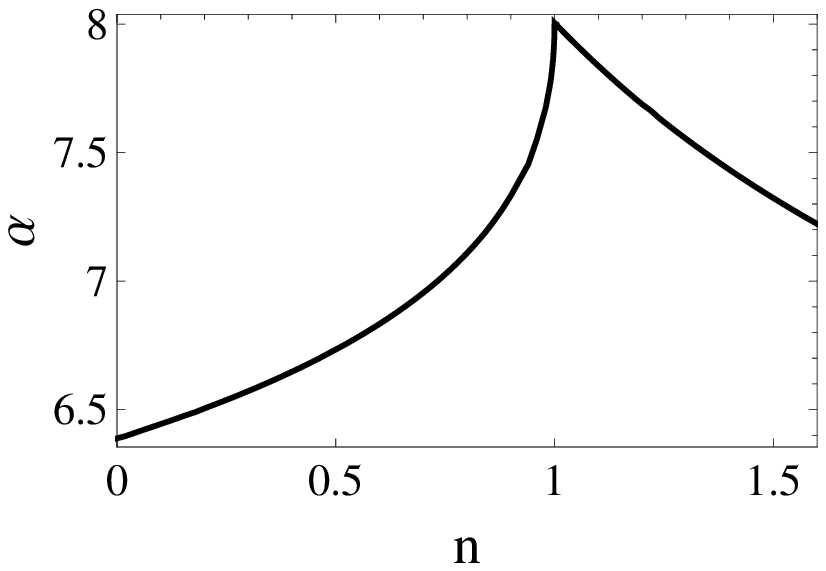}
\put(18,38){\includegraphics[scale=.35]{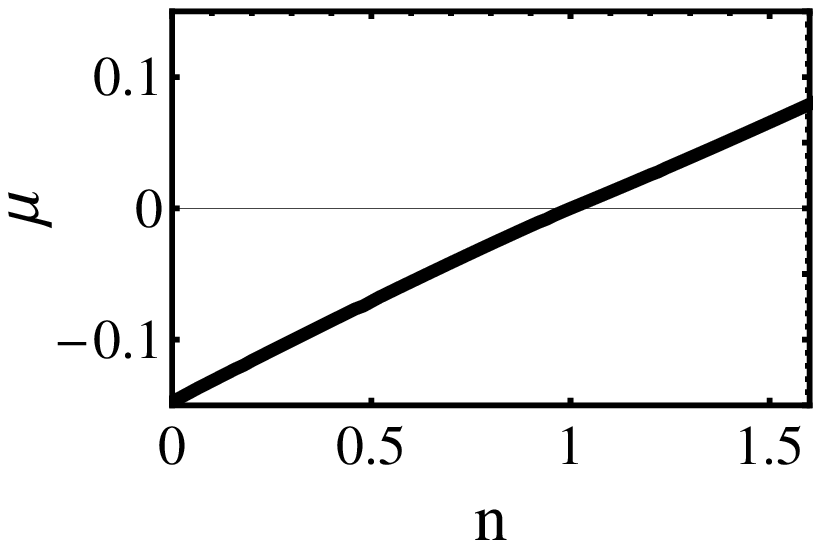}}
\end{overpic}
\caption{The second derivative of the thermodynamic potential $\Omega(\mu)$ with respect to $\mu$ as a function of the density $n$ for the isotropic $p_x+ip_y$ pairing (upper panel) and the anisotropic $p_x$ pairing (lower panel). The inserts show the chemical potential $\mu$ as a function of $n$. In this calculation, we employ the NSR-type potential~\cite{BCSBEC01} for the 2D case, 
$\Gamma({\bf k})=(k/k_1)g(\varphi)/(1+k/k_0)^{3/2}$~\cite{QPT01}. Here $g(\varphi)=e^{i\varphi}$ for $p_x+ip_y$ pairing and $g(\varphi)=\cos\varphi$ for $p_x$ pairing. In this potential model, $k_0 \sim r_0^{-1}$, where $r_0$ plays the role of the interaction range. $k_0$ and $k_1$ set the momentum scales in the short and long wavelength limits, respectively. The density $n$ has been normalized such that $n=1$ at $\mu=0$. All quantities are shown in proper units. \label{fig1}}
\end{figure}

In Fig. \ref{fig1}, we show the numerical results of $\alpha(\mu)$ for two typical $p$-wave pairings: the complex and isotropic $p_x+ip_y$ pairing and the anisotropic $p_x$ pairing. For the $p_x+ip_y$ pairing, the superfluid state is fully gapped for both $\mu>0$ and $\mu<0$, and it is only gapless at $\mu=0$. For $p_x$ pairing, the superfluid state is gapped for $\mu<0$ and gapless for $\mu>0$. In the calculation we have adopted a NSR-type potential to regularize the ultraviolet divergence in the gap equation. However, as we have argued above, the nonanalytical behavior at $\mu=0$ does not depend on this specific choice. The results in Fig. \ref{fig1} show that the nonanalytical behavior appears for both isotropic and anisotropic pairings, consistent with our conclusion that the nonanalytical behavior is solely due to the infrared behavior of the pairing interaction.

\subsection{Collective modes}

To study the behavior of the collective modes across the BCS-BEC quantum phase transition, we consider the Gaussian-fluctuation part  ${\cal S}_{\text{eff}}^{(2)}[\phi,\phi^*]$. It can be written in the bilinear form
\begin{equation}
{\cal S}_{\text{eff}}^{(2)}=\frac{1}{2}\sum_Q \Lambda^\dagger(Q) {\bf M}(Q) \Lambda(Q),
\end{equation}
where $Q=(i\nu_n,{\bf q})$ with $\nu_n$ being the boson Matsubara frequency, $\Lambda(Q)=[\phi(Q), \phi^*(-Q)]^{\rm T}$, and the $2\times2$ matrix ${\bf M}(Q)$ is the inverse of the collective-mode propagator. The matrix elements of ${\bf M}(Q)$ are constructed by using the mean-field fermion propagator ${\cal G}(K)$ which is obtained from the expression of ${\bf G}_{{\bf k},{\bf k}^\prime}(\tau)$. Here $K=(i\omega_n,{\bf k})$, with $\omega_n$ being the fermion Matsubara frequency. We have
\begin{eqnarray}
{\cal G}^{-1}(K)=\left(\begin{array}{cc}i\omega_n-\xi_{\bf k} & \Delta_0\Gamma_p({\bf k})\\
\Delta_0\Gamma_p^*({\bf k})& i\omega_n+\xi_{\bf k}\end{array}\right).
\end{eqnarray}
Then the matrix elements of ${\bf M}(Q)$ can be expressed as
\begin{eqnarray}
{\bf M}_{11}(Q)&=&\frac{1}{\lambda_p}+\frac{1}{4}\sum_K|\Gamma_p({\bf k})|^2\text{Tr}\left[{\cal G}_{11}(K_+){\cal G}_{22}(K_-)\right],\nonumber\\
{\bf M}_{12}(Q)&=&\frac{1}{4}\sum_K|\Gamma_p({\bf k})|^2\text{Tr}\left[{\cal G}_{12}(K_+){\cal G}_{12}(K_-)\right],
\end{eqnarray}
where $K_\pm=K\pm Q/2$. In addition, we have ${\bf M}_{22}(Q)={\bf M}_{11}(-Q)$ and ${\bf M}_{21}(Q)={\bf M}_{12}(Q)$. Completing the summation over the boson Matsubara frequency, we obtain at zero temperature
\begin{eqnarray}
&&{\bf M}_{11}(Q)\nonumber\\
&=&\frac{1}{\lambda_p}+\sum_{\bf k}\frac{|\Gamma_p({\bf k})|^2}{2}
\left(\frac{u_+^2u_-^2}{i\nu_n-E_+-E_-}-\frac{\upsilon_+^2\upsilon_-^2}{i\nu_n+E_++E_-}\right),\nonumber\\
&&{\bf M}_{12}(Q)\nonumber\\
&=&-\sum_{\bf k}\frac{|\Gamma_p({\bf k})|^2}{2}
\left(\frac{u_+\upsilon_+u_-\upsilon_-}{i\nu_n-E_+-E_-}-\frac{u_+\upsilon_+u_-\upsilon_-}{i\nu_n+E_++E_-}\right).
\end{eqnarray}
Here $+$ and $-$ denote the momenta ${\bf k}+{\bf q}/2$ and ${\bf k}-{\bf q}/2$, respectively. $u_{\bf k}^2$ and $\upsilon_{\bf k}^2$ are the
BCS distribution functions defined as $u_{\bf k}^2=1-\upsilon_{\bf k}^2=(1/2)(1+\xi_{\bf k}/E_{\bf k})$.

Taking the analytical continuation $i\nu_n\rightarrow\omega+i0^+$, the dispersions $\omega({\bf q})$ of the collective modes are
determined by the equation $\det{{\bf M}[\omega({\bf q}), {\bf q}]}=0$. It is usual to decompose ${\bf M}_{11}(\omega,{\bf q})$ as ${\bf M}_{11}(\omega,{\bf q})={\bf M}_{11}^+(\omega,{\bf q})+{\bf M}_{11}^-(\omega,{\bf q})$, where ${\bf M}_{11}^+(\omega,{\bf q})$ and ${\bf M}_{11}^-(\omega,{\bf q})$ are even and odd functions of $\omega$, respectively. Meanwhile ${\bf M}_{12}(\omega,{\bf q})$ and ${\bf M}_{21}(\omega,{\bf q})$ are even functions of $\omega$ automatically. Then we separate the complex field $\phi(Q)$ into
its amplitude part $\lambda(Q)$ and phase part $\theta(Q)$, $\phi(Q)=\lambda(Q)+i\Delta_0\theta(Q)$. The effective action ${\cal S}_{\text{eff}}^{(2)}$ becomes
\begin{equation}
{\cal S}_{\text{eff}}^{(2)}=\frac{1}{2}\sum_Q\left(\begin{array}{cc} \lambda^*(Q)&\theta^*(Q)\end{array}\right){\bf N}(Q)\left(\begin{array}{c} \lambda(Q)\\ \theta(Q)\end{array}\right),
\end{equation}
where the matrix elements of ${\bf N}(Q)$ read ${\bf N}_{11}(Q)=2({\bf M}_{11}^++{\bf M}_{12})$, ${\bf N}_{22}(Q)=2\Delta_0^2({\bf M}_{11}^+
-{\bf M}_{12})$, ${\bf N}_{12}(Q)=2i\Delta_0{\bf M}_{11}^-$, and ${\bf N}_{21}(Q)=-2i\Delta_0{\bf M}_{11}^-$. Since
${\bf M}_{11}^-(0,{\bf q})=0$, the amplitude and phase modes decouple completely at $\omega=0$. At the saddle point we have precisely ${\bf M}_{11}^+(0,{\bf 0})={\bf M}_{12}(0,{\bf 0})$. Therefore, the phase mode at ${\bf q}=0$ is gapless, that is, the Goldstone mode or the Anderson-Bogoliubov mode for fermionic superfluids.

To study the low-energy behavior of the collective modes, we make a small ${\bf q}$ and $\omega$ expansion of ${\bf N}(Q)$ at zero temperature. In general, the expansion takes the form ${\bf N}_{11}= A+C{\bf q}^2-D\omega^2+\cdots$, ${\bf N}_{22}= J{\bf q}^2-R\omega^2+\cdots$, and
${\bf N}_{12}=-{\bf N}_{21}=-iB\omega+\cdots$. The explicit forms of the expansion parameters $A,B,D$, and $R$ can be evaluated as
\begin{eqnarray}
&&A=\frac{\Delta_0^2}{2}\sum_{\bf k}\frac{|\Gamma_p({\bf k})|^4}{E_{\bf k}^3},\nonumber\\
&&B=\frac{\Delta_0}{4}\sum_{\bf k}\frac{\xi_{\bf k}|\Gamma_p({\bf k})|^2}{E_{\bf k}^3},\nonumber\\
&&D=\frac{1}{8}\sum_{\bf k}\frac{\xi_{\bf k}^2|\Gamma_p({\bf k})|^2}{E_{\bf k}^5},\nonumber\\
&&R=\frac{\Delta_0^2}{8}\sum_{\bf k}\frac{|\Gamma_p({\bf k})|^2}{E_{\bf k}^3}.
\end{eqnarray}
The phase stiffness $J$ is related to the superfluid density $n_s$ by $J=n_s/(4M)$. The superfluid density $n_s$ can also be obtained from its standard definition \cite{NS01}. When the superfluid moves with a uniform velocity $\mbox{\boldmath{$\upsilon$}}_s$, the pair field transforms as
$\Phi\rightarrow \Phi e^{2iM\mbox{\boldmath{$\upsilon$}}_s\cdot {\bf r}}$. The superfluid density $n_s$ is defined as the response of the thermodynamic potential $\Omega$ to an infinitesimal velocity $\mbox{\boldmath{$\upsilon$}}_s$; that is,
$\Omega(\mbox{\boldmath{$\upsilon$}}_s)=\Omega({\bf 0})+\frac{1}{2}n_s\mbox{\boldmath{$\upsilon$}}_s^2+O(\mbox{\boldmath{$\upsilon$}}_s^4)$.
For the present system, the superfluid density equals the total fermion density $n$ at zero temperature, guaranteed by the Galilean invariance.

\begin{figure}[!htb]
\begin{center}
\includegraphics[width=8cm]{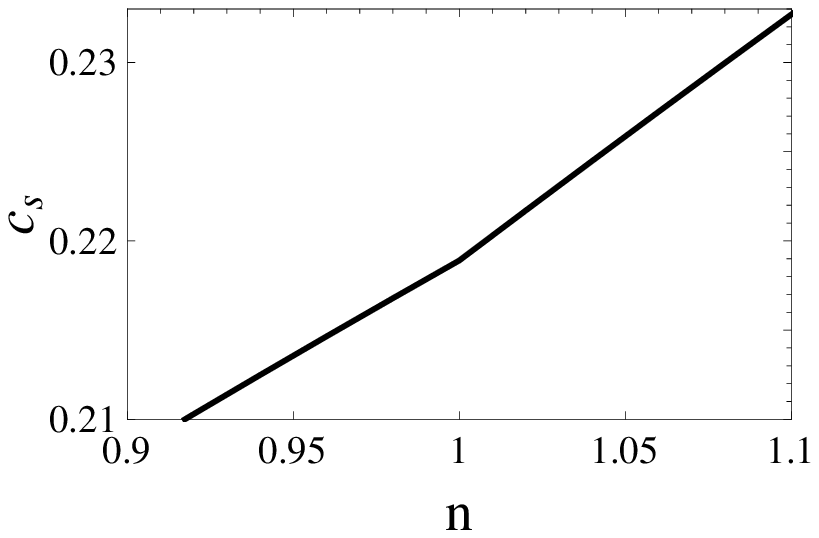}
\includegraphics[width=8cm]{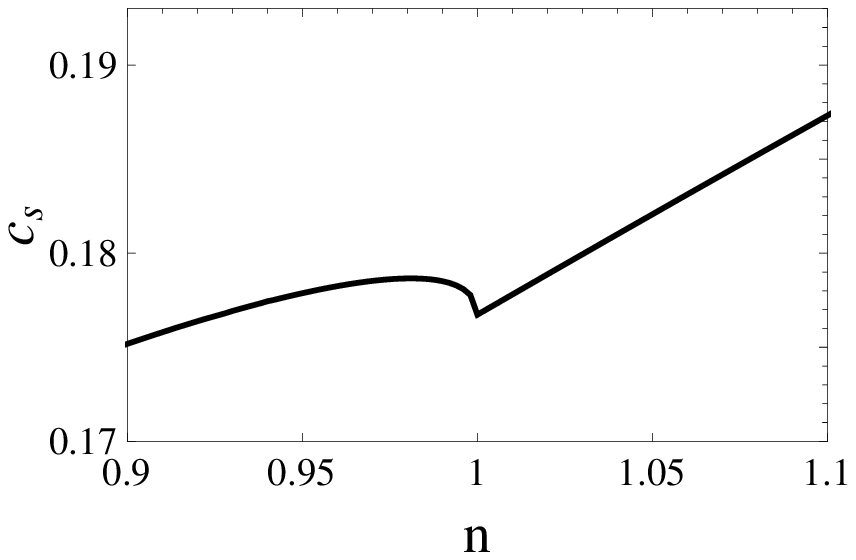}
\caption{The velocity of the Anderson-Bogoliubov mode as a function of the density $n$ for the isotropic $p_x+ip_y$ pairing (upper panel) and the anisotropic $p_x$ pairing (lower panel). In the calculation we employ the same potential model as used in Fig. \ref{fig1}. All quantities are shown in proper units.
 \label{fig2}}
\end{center}
\end{figure}

The dispersion of the gapless Anderson-Bogoliubov mode is given by
\begin{eqnarray}
\omega({\bf q})=c_s|{\bf q}|,
\end{eqnarray}
where the sound velocity $c_s$ reads
\begin{eqnarray}
c_s=\sqrt{\frac{J}{R+B^2/A}}.
\end{eqnarray}
By comparing the expansion parameters $A,B$, and $R$ with the expressions in Eq. (21), we find that
\begin{eqnarray}
&&R=\frac{1}{4}\frac{\partial n(\mu,\Delta_0)}{\partial\mu},\nonumber\\
&&B=\frac{1}{2}\frac{\partial n(\mu,\Delta_0)}{\partial\Delta_0},\nonumber\\
&&A=\frac{\partial^2{\cal V}(\Delta_0,\mu)}{\partial\Delta_0^2}.
\end{eqnarray}
Therefore, we have
\begin{eqnarray}
c_s=\sqrt{\frac{n}{M\alpha}}.
\end{eqnarray}
We note that this relation is quite generic. It also applies to $s$-wave fermionic superfluids. Since $n$ goes smoothly with $\mu$ we conclude that the sound velocity behaves nonanalytically across the quantum phase transition point $\mu=0$. In Fig. \ref{fig2}, we show the behavior of the sound
velocity $c_s$ around the BCS-BEC quantum phase transition, using the same potential model as employed in Fig. \ref{fig1}. It shows obviously that the evolution of the Anderson-Bogoliubov mode is not smooth, corresponding to the BCS-BEC quantum phase transition.

\section{$d$-wave pairing in spin-$\frac{1}{2}$ Fermi gases}
The many-body Hamiltonian of 2D spin-$\frac{1}{2}$ fermions can be written as $H=H_0+H_{\rm int}$, where the single-particle part reads $H_0=\sum_{\bf k}\xi_{\bf k}(\psi^{\dagger}_{\uparrow,{\bf k}}\psi^{\phantom{\dag}}_{\uparrow,{\bf k}}+\psi^{\dagger}_{\downarrow,{\bf k}}\psi^{\phantom{\dag}}_{\downarrow,{\bf k}})$ and the $d$-wave interaction part is
\begin{eqnarray}
H_{\rm int}=\sum_{{\bf k},{\bf k}^\prime,{\bf q}}V_d({\bf k},{\bf k}^\prime)\psi^\dagger_{\uparrow,{\bf k}+{\bf q}/2}\psi^\dagger_{\downarrow,-{\bf k}+{\bf q}/2}\psi^{\phantom{\dag}}_{\downarrow,-{\bf k}^\prime+{\bf q}/2}\psi^{\phantom{\dag}}_{\uparrow,{\bf k}^\prime+{\bf q}/2}.
\end{eqnarray}
The generic infrared behavior of the $d$-wave interaction potential is $V_d({\bf k},{\bf k}^\prime)\sim k^2(k^\prime)^2$ for $k,k^\prime\rightarrow 0$. Here we also consider a separable potential for $d$-wave interaction, $V_d({\bf k},{\bf k}^\prime)=-\lambda_d\Gamma_d({\bf k})\Gamma_d^*({\bf k}^\prime)$, where $\lambda_d$ is the $d$-wave coupling constant and the function $\Gamma_d(\bf k)$ characterizes the $d$-wave pairing symmetry. The infrared behavior of the function $\Gamma_d(\bf k)$ is
\begin{eqnarray}
\Gamma_d({\bf k})\sim k^2,\ \ \ \ \ {\bf k}\rightarrow 0.
\end{eqnarray}
As a result, the infrared divergence at $\mu=0$ is more pronounced for $d$-wave pairing.

The partition function of the system is given by ${\cal Z}=\int \mathcal{ D} \psi \mathcal{D}\psi^\dagger\exp\left\{-{\cal S}[\psi,\psi^\dagger]\right\}$, where the action reads
\begin{eqnarray}
{\cal S}[\psi,\psi^\dagger]=\int_0^{1/T} d\tau\left[\sum_{{\bf k},\sigma}\psi_{\sigma,{\bf k}}^\dagger(\tau)\partial_\tau
\psi^{\phantom{\dag}}_{\sigma,{\bf k}}(\tau)+ H(\psi,\psi^\dagger)\right].
\end{eqnarray}
Again, we introduce the auxiliary complex pairing field $\Phi_{\bf q}(\tau)$, which couples to $\psi_\uparrow^\dagger\psi_\downarrow^\dagger$, and apply the Hubbard-Stratonovich transformation. With the help of the Nambu-Gor$'$kov basis defined as
$\Psi_{\bf k}=(\psi^{\phantom{\dag}}_{\uparrow, {\bf k}},\psi^\dagger_{\downarrow, {\bf k}})^{\rm T}$ we obtain $\mathcal {Z}=\int\mathcal{D} \Phi \mathcal{D} \Phi^{\ast}\exp\big\{-{\cal S}_{\rm{eff}}[\Phi, \Phi^{\ast}]\big\}$, where
\begin{equation}
{\cal S}_{\rm{eff}} = \int_0^{1/T}d\tau \left[\sum_{\bf q}\frac{|\Phi_{\bf q}(\tau)|^{2}}{\lambda_d}
 +\sum_{\bf k,k^\prime}\left(\xi_{\bf k}\delta_{\bf k,k^\prime}- \mbox{Trln}{\bf G}_{\bf k,k^\prime}^{-1}\right)\right].
\end{equation}
Here the inverse single-particle Green's function ${\bf G}_{\bf k,k^\prime}^{-1}$ takes a similar form:
\begin{eqnarray}
{\bf G}_{\bf k,k^\prime}^{-1}=\left(\begin{array}{cc}-(\partial_{\tau}+\xi_{\bf k})\delta_{\bf k,k^\prime} & \Phi_{{\bf k}-{\bf k}^\prime}(\tau)\Gamma_d(\frac{{\bf k}+{\bf k}^\prime}{2})\\
\Phi^*_{{\bf k}-{\bf k}^\prime}(\tau)\Gamma_d^*(\frac{{\bf k}+{\bf k}^\prime}{2})& -(\partial_{\tau}-\xi_{\bf k})\delta_{\bf k,k^\prime}\end{array}\right).
\end{eqnarray}

Following the same method used in the previous section, we decompose the pairing field as $\Phi_{\bf q}(\tau)=\Delta_0\delta_{\bf q,0}+\phi_{\bf q}(\tau)$, where $\phi$ is the fluctuation around the mean field. The effective action ${\cal S}_{\rm{eff}}[\Phi,\Phi^*]$ can be expanded in powers of the fluctuation $\phi$,
${\cal S}_{\rm{eff}}[\Phi,\Phi^*]={\cal S}_{\rm{eff}}^{(0)}(\Delta_0)+{\cal S}_{\rm{eff}}^{(2)}[\phi,\phi^*]+\cdots$,
where ${\cal S}_{\rm{eff}}^{(0)}(\Delta_0)$ is the saddle point or mean-field effective action with $\Delta_0$
determined by the saddle point condition $\partial{\cal S}_{\rm{eff}}^{(0)}/\partial\Delta_0=0$.

\subsection{Quantum phase transition}
The effective potential for $d$-wave pairing at zero temperature is given by
\begin{eqnarray}
{\cal V}(\Delta_0,\mu)=\frac{\Delta_0^2}{\lambda_d}+\sum_{\bf k}\left(\xi_{\bf k}-E_{\bf k}\right).
\end{eqnarray}
Note that the factor $1/2$ in Eq. (10) is replaced by $1$ here due to the appearance of spin degree of freedom. The single-particle excitation spectrum $E_{\bf k}$ is also given by
\begin{eqnarray}
E_{\bf k}=\sqrt{\xi_{\bf k}^2+|\Delta({\bf k})|^2},
\end{eqnarray}
where $\Delta({\bf k})=\Delta_0\Gamma_d({\bf k})$ is the gap function for $d$-wave pairing. The order parameter $\Delta_0$ is determined by
the $d$-wave gap equation
\begin{eqnarray}
\frac{1}{\lambda_d}=\sum_{\bf k}\frac{|\Gamma_d({\bf k})|^2}{2E_{\bf k}}.
\end{eqnarray}

To study the nature of the quantum phase transition, we also focus on the thermodynamic potential $\Omega(\mu)\equiv {\cal V}(\Delta_0(\mu),\mu)$.
The first derivative of the thermodynamic potential with respect to $\mu$ gives the number equation,
\begin{eqnarray}
n(\mu,\Delta_0)=-\frac{\partial \Omega(\mu)}{\partial\mu}=\sum_{\bf k}\left(1-\frac{\xi_{\bf k}}{E_{\bf k}}\right).
\end{eqnarray}
Then we consider the second derivative
\begin{eqnarray}
\alpha(\mu)=-\frac{\partial^2\Omega(\mu)}{\partial\mu^2}.
\end{eqnarray}
Again, this quantity can be expressed as
\begin{eqnarray}
\alpha(\mu)=\frac{\partial n(\mu,\Delta_0)}{\partial\mu}+\left(\frac{\partial n(\mu,\Delta_0)}{\partial\Delta_0}\right)^2
\left(\frac{\partial^2{\cal V}(\Delta_0,\mu)}{\partial\Delta_0^2}\right)^{-1}.
\end{eqnarray}
We note that the second term was missing in the previous study~\cite{QPT02}. However, the divergent behavior of $\alpha$ is governed by the first term. Therefore, the results in Ref. \cite{QPT02} are still reliable. The derivatives in Eq. (44) can be explicitly
evaluated as
\begin{eqnarray}
&&\frac{\partial n(\mu,\Delta_0)}{\partial\mu}=\Delta_0^2\sum_{\bf k}\frac{|\Gamma_d({\bf k})|^2}{E_{\bf k}^3},\nonumber\\
&&\frac{\partial n(\mu,\Delta_0)}{\partial\Delta_0}=\Delta_0\sum_{\bf k}\frac{\xi_{\bf k}|\Gamma_d({\bf k})|^2}{E_{\bf k}^3},\nonumber\\
&&\frac{\partial^2{\cal V}(\Delta_0,\mu)}{\partial\Delta_0^2}=\Delta_0^2\sum_{\bf k}\frac{|\Gamma_d({\bf k})|^4}{E_{\bf k}^3}.
\end{eqnarray}

\begin{figure}
\centering
\begin{overpic}
[scale=.92]{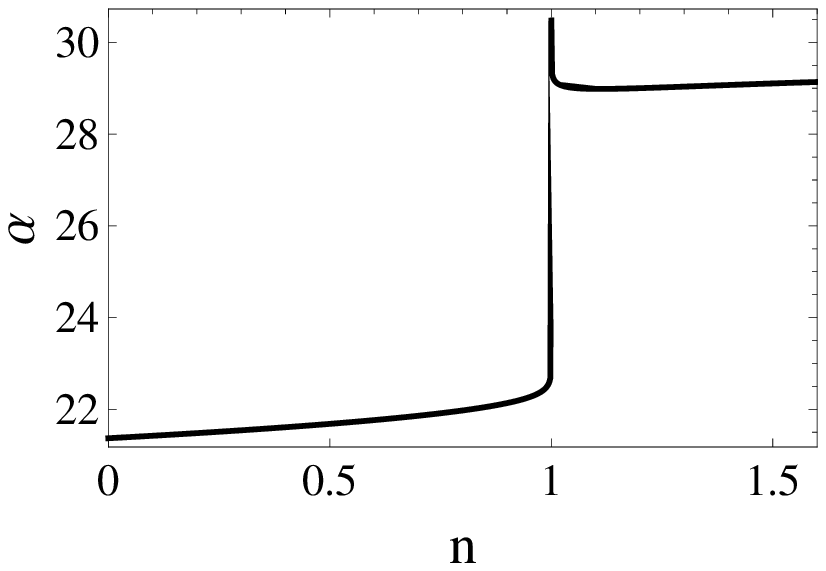}
\put(16,35){\includegraphics[scale=.39]{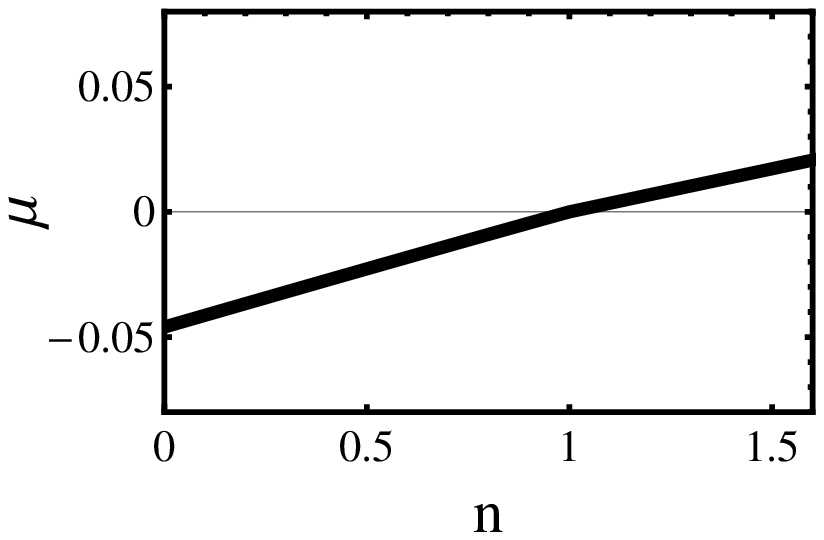}}
\end{overpic}
\begin{overpic}
[scale=.92]{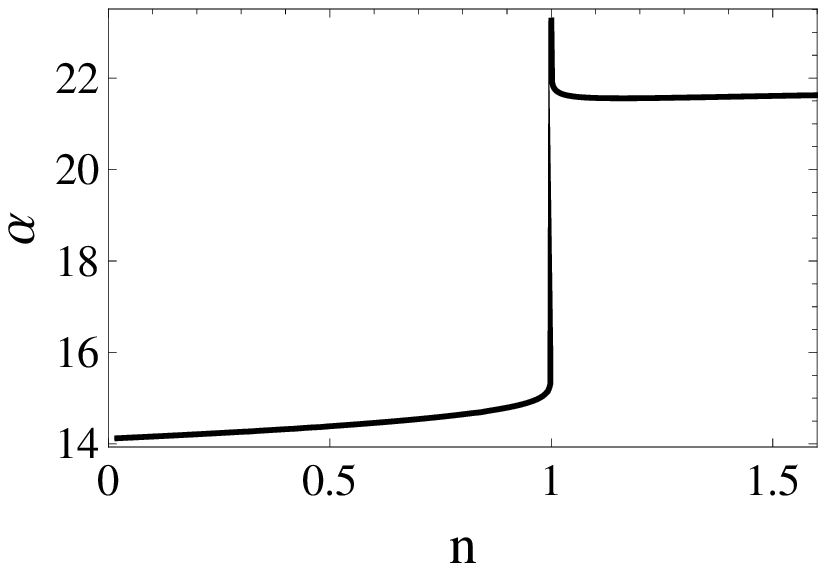}
\put(16,35){\includegraphics[scale=.39]{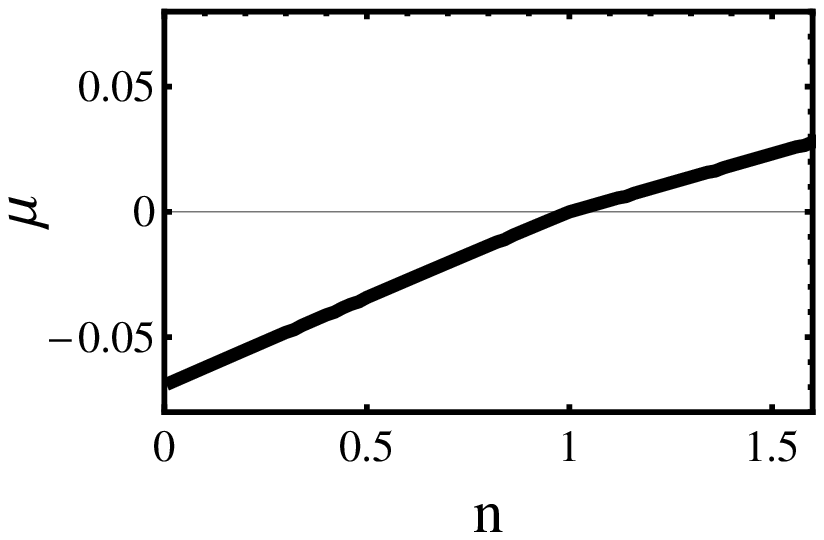}}
\end{overpic}
\caption{The second derivative of the thermodynamic potential $\Omega(\mu)$ with respect to $\mu$ as a function of the density $n$ for the isotropic $d_{x^2-y^2}+2id_{xy}$ pairing (upper panel) and the anisotropic $d_{x^2-y^2}$ pairing (lower panel). The inserts show the chemical potential $\mu$ as a function of $n$. In this calculation, we employ the NSR-type potential for $d$-wave interaction,
$\Gamma({\bf k})=(k/k_1)^2g(\varphi)/(1+k/k_0)^{5/2}$~\cite{QPT02}. Here  $g(\varphi)=\cos2\varphi$ for $d_{x^2-y^2}$ pairing and $g(\varphi)=e^{2i\varphi}$ for $d_{x^2-y^2}+2id_{xy}$ pairing. The density $n$ has been normalized such that $n=1$ at $\mu=0$. All quantities are shown in proper units. \label{fig3}}
\end{figure}

By using a NSR-type potential, Botelho and Sa de Melo showed that $\alpha$ is divergent at $\mu=0$ \cite{QPT02}. Here we show that this divergence generally appears due to an infrared divergence at $\mu=0$. For ${\bf k}\rightarrow 0$, we have $\Gamma_d({\bf k})\sim k^2$ and therefore $E_{\bf k}\sim k^2$ at $\mu=0$. As a result, the momentum integral
\begin{eqnarray}
{\cal I}_2(\mu)=\sum_{\bf k}\frac{|\Gamma_d({\bf k})|^2}{E_{\bf k}^3}
\end{eqnarray}
is infrared divergent at $\mu=0$. Actually, at $\mu=0$, the infrared behavior of the above integral is
\begin{eqnarray}
{\cal I}_2(0)\sim\int_0^\epsilon kdk\frac{k^4}{k^6}\sim \int_0^\epsilon \frac{dk}{k}.
\end{eqnarray}
Therefore, the quantity $\alpha(\mu)$ itself is divergent at $\mu=0$, which indicates a quantum phase transition. This is not surprising, because the infrared divergence should be more pronounced for $d$-wave pairing.

The above discussion shows that a quantum phase transition occurs at $\mu=0$ for $d$-wave pairing, indicated by the divergence of the quantity $\alpha(\mu)$. The divergence is also caused solely by the infrared behavior of the $d$-wave pairing potential and is independent of the details of the interaction as well as the order parameter symmetry (i.e., the angle dependence of the gap function).

In Fig. \ref{fig3}, we show the numerical results of $\alpha(\mu)$ for two typical $d$-wave pairings: the complex and isotropic $d_{x^2-y^2}+2id_{xy}$ pairing and the anisotropic $d_{x^2-y^2}$ pairing. For $d_{x^2-y^2}+2id_{xy}$ pairing, the superfluid state is fully gapped for both $\mu>0$ and $\mu<0$, and it is only gapless at $\mu=0$. For $d_{x^2-y^2}$ pairing, the superfluid state is gapped for $\mu<0$ and gapless for $\mu>0$. In the calculation we have also adopted a NSR-type potential to regularize the ultraviolet divergence in the gap equation. However, as we mentioned above, the divergence at $\mu=0$ does not depend on this specific choice. The results in Fig. \ref{fig3} show that the divergence appears for both isotropic and anisotropic pairings, consistent with our conclusion that the divergence is solely due to the infrared behavior of the $d$-wave pairing interaction.

\subsection{Collective modes}
To study the behavior of the collective modes across the BCS-BEC quantum phase transition for $d$-wave pairing, we also consider the Gaussian-fluctuation part ${\cal S}_{\text{eff}}^{(2)}[\phi,\phi^*]$, which can also be written as
\begin{equation}
{\cal S}_{\text{eff}}^{(2)}=\frac{1}{2}\sum_Q \Lambda^\dagger(Q) {\bf M}(Q) \Lambda(Q).
\end{equation}
The matrix elements of ${\bf M}(Q)$ are also constructed using the mean-field fermion propagator ${\cal G}(K)$ given by
\begin{eqnarray}
{\cal G}^{-1}(K)=\left(\begin{array}{cc}i\omega_n-\xi_{\bf k} & \Delta_0\Gamma_d({\bf k})\\
\Delta_0\Gamma_d^*({\bf k})& i\omega_n+\xi_{\bf k}\end{array}\right).
\end{eqnarray}
We have
\begin{eqnarray}
{\bf M}_{11}(Q)&=&\frac{1}{\lambda_d}+\frac{1}{2}\sum_K|\Gamma_d({\bf k})|^2\text{Tr}\left[{\cal G}_{11}(K_+){\cal G}_{22}(K_-)\right],\nonumber\\
{\bf M}_{12}(Q)&=&\frac{1}{2}\sum_K|\Gamma_d({\bf k})|^2\text{Tr}\left[{\cal G}_{12}(K_+){\cal G}_{12}(K_-)\right].
\end{eqnarray}
The relations ${\bf M}_{22}(Q)={\bf M}_{11}(-Q)$ and ${\bf M}_{21}(Q)={\bf M}_{12}(Q)$ also hold here. By completing the summation over the boson Matsubara frequency, we obtain at zero temperature
\begin{eqnarray}
&&{\bf M}_{11}(Q)\nonumber\\
&=&\frac{1}{\lambda_d}+\sum_{\bf k}|\Gamma_d({\bf k})|^2
\left(\frac{u_+^2u_-^2}{i\nu_n-E_+-E_-}-\frac{\upsilon_+^2\upsilon_-^2}{i\nu_n+E_++E_-}\right),\nonumber\\
&&{\bf M}_{12}(Q)\nonumber\\
&=&-\sum_{\bf k}|\Gamma_d({\bf k})|^2
\left(\frac{u_+\upsilon_+u_-\upsilon_-}{i\nu_n-E_+-E_-}-\frac{u_+\upsilon_+u_-\upsilon_-}{i\nu_n+E_++E_-}\right).
\end{eqnarray}
Here the notations are the same as in the last section.

The dispersions $\omega({\bf q})$ of the collective modes are also determined by the equation $\det{{\bf M}[\omega({\bf q}), {\bf q}]}=0$. Again, we decompose ${\bf M}_{11}(\omega,{\bf q})$ as ${\bf M}_{11}(\omega,{\bf q})={\bf M}_{11}^+(\omega,{\bf q})+{\bf M}_{11}^-(\omega,{\bf q})$, where ${\bf M}_{11}^+(\omega,{\bf q})$ and ${\bf M}_{11}^-(\omega,{\bf q})$ are even and odd functions of $\omega$, respectively. Then we separate the complex field $\phi(Q)$ into
its amplitude part $\lambda(Q)$ and phase part $\theta(Q)$, $\phi(Q)=\lambda(Q)+i\Delta_0\theta(Q)$. The effective action ${\cal S}_{\text{eff}}^{(2)}$ becomes
\begin{equation}
{\cal S}_{\text{eff}}^{(2)}=\frac{1}{2}\sum_Q\left(\begin{array}{cc} \lambda^*(Q)&\theta^*(Q)\end{array}\right){\bf N}(Q)\left(\begin{array}{c} \lambda(Q)\\ \theta(Q)\end{array}\right),
\end{equation}
where the matrix elements of ${\bf N}(Q)$ read ${\bf N}_{11}(Q)=2({\bf M}_{11}^++{\bf M}_{12})$, ${\bf N}_{22}(Q)=2\Delta_0^2({\bf M}_{11}^+
-{\bf M}_{12})$, ${\bf N}_{12}(Q)=2i\Delta_0{\bf M}_{11}^-$, and ${\bf N}_{21}(Q)=-2i\Delta_0{\bf M}_{11}^-$. At the saddle point we also have precisely ${\bf M}_{11}^+(0,{\bf 0})={\bf M}_{12}(0,{\bf 0})$ for $d$-wave pairing. Therefore, the phase mode at ${\bf q}=0$ is gapless,  corresponding to the Anderson-Bogoliubov mode for $d$-wave pairing.

The small ${\bf q}$ and $\omega$ expansion of ${\bf N}(Q)$ also takes the form ${\bf N}_{11}= A+C{\bf q}^2-D\omega^2+\cdots$, ${\bf N}_{22}= J{\bf q}^2-R\omega^2+\cdots$, and ${\bf N}_{12}=-{\bf N}_{21}=-iB\omega+\cdots$. The explicit forms of the expansion parameters $A,B,D$, and $R$ here are given by
\begin{eqnarray}
&&A=\Delta_0^2\sum_{\bf k}\frac{|\Gamma_d({\bf k})|^4}{E_{\bf k}^3},\nonumber\\
&&B=\frac{\Delta_0}{2}\sum_{\bf k}\frac{\xi_{\bf k}|\Gamma_d({\bf k})|^2}{E_{\bf k}^3},\nonumber\\
&&D=\frac{1}{4}\sum_{\bf k}\frac{\xi_{\bf k}^2|\Gamma_d({\bf k})|^2}{E_{\bf k}^5},\nonumber\\
&&R=\frac{\Delta_0^2}{4}\sum_{\bf k}\frac{|\Gamma_d({\bf k})|^2}{E_{\bf k}^3}.
\end{eqnarray}
The phase stiffness $J$ is also related to the superfluid density $n_s$ by $J=n_s/(4M)$ with $n_s=n$ at zero temperature.

\begin{figure}[!htb]
\begin{center}
\includegraphics[width=8cm]{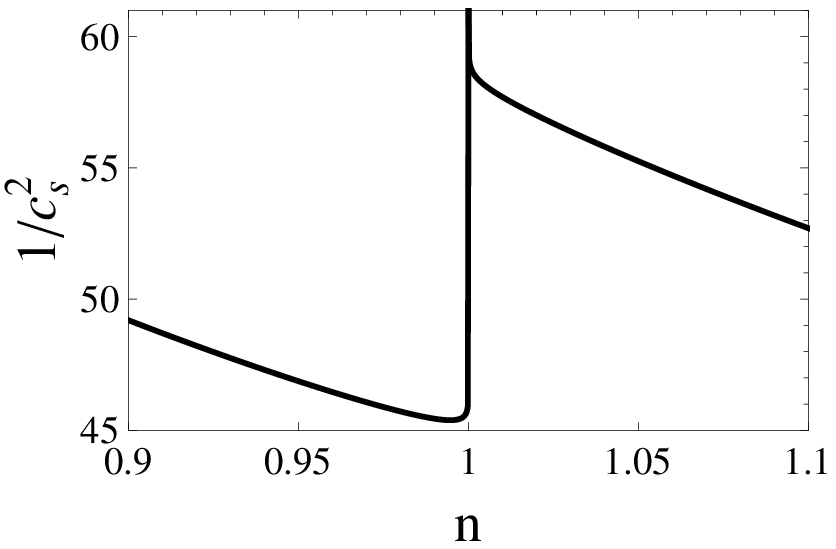}
\includegraphics[width=8cm]{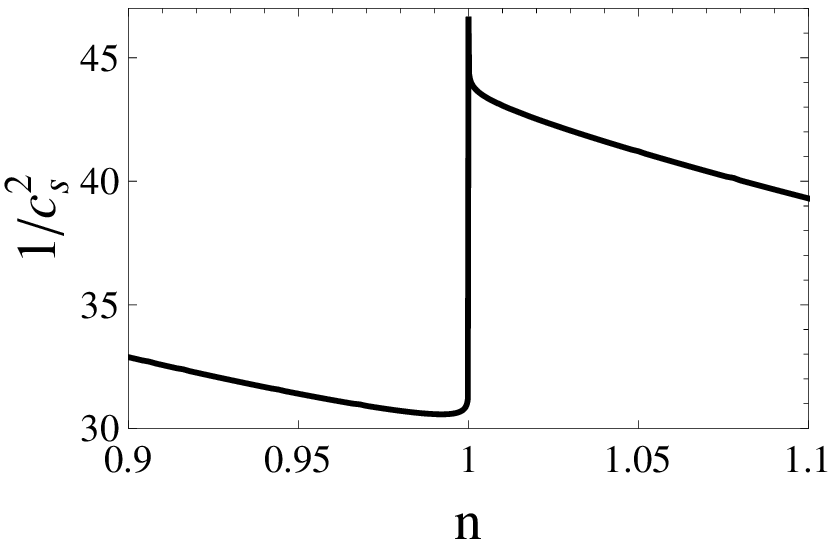}
\caption{The velocity of the Anderson-Bogoliubov mode as a function of the density $n$ for the isotropic $d_{x^2-y^2}+2id_{xy}$ pairing (upper panel) and the anisotropic $d_{x^2-y^2}$ pairing (lower panel). In the calculation we employ the same potential model as used in Fig. \ref{fig3}. All quantities are shown in proper units.
 \label{fig4}}
\end{center}
\end{figure}

The dispersion of the gapless Anderson-Bogoliubov mode is given by $\omega({\bf q})=c_s|{\bf q}|$,
where the sound velocity $c_s$ reads
\begin{eqnarray}
c_s=\sqrt{\frac{J}{R+B^2/A}}.
\end{eqnarray}
Comparing the expansion parameters $A,B$, and $R$ with the expressions in Eq. (45), we also obtain the following relations for $d$-wave pairing:
\begin{eqnarray}
&&R=\frac{1}{4}\frac{\partial n(\mu,\Delta_0)}{\partial\mu},\nonumber\\
&&B=\frac{1}{2}\frac{\partial n(\mu,\Delta_0)}{\partial\Delta_0},\nonumber\\
&&A=\frac{\partial^2{\cal V}(\Delta_0,\mu)}{\partial\Delta_0^2}.
\end{eqnarray}
Therefore, for $d$-wave pairing we also have
\begin{eqnarray}
c_s=\sqrt{\frac{n}{M\alpha}}.
\end{eqnarray}
Since the quantity $\alpha$ is divergent at the quantum phase transition point $\mu=0$, the sound velocity $c_s$ goes nonanalytically across the quantum phase transition and vanishes at $\mu=0$. In Fig. \ref{fig4}, we show the behavior of the sound
velocity $c_s$ around the BCS-BEC quantum phase transition, using the same potential model for $d$-wave pairing as employed in Fig. \ref{fig3}. It shows that the evolution of the Anderson-Bogoliubov mode for $d$ wave is also not smooth. The vanishing of the sound velocity $c_s$ may bring interesting thermodynamic consequences. For instance, the low-temperature specific heat caused by the Goldstone mode can be given by
\begin{eqnarray}
C_{\rm v}=\frac{2\pi^2}{15c_s^3}T^4.
\end{eqnarray}
Therefore, the low-temperature specific heat of the fermionic superfluids near $\mu=0$ should be very large for $d$-wave pairing.

On the other hand, the expansion parameter $D$, which is related to the massive amplitude mode or the so-called Anderson-Higgs mode, is divergent at the quantum phase transition point $\mu=0$. A rough estimation of the mass gap of the amplitude mode can be given by $M_{\rm{AH}}=\sqrt{(A+B^2/R)/D}$, which vanishes at $\mu=0$. This indicates that the amplitude mode gets softened around the quantum phase transition point, which could be interesting for future identification of the massive Anderson-Higgs mode in fermionic superfluids.

\section{Summary}
While it is generally accepted that the BCS-BEC evolution in fermionic systems with $s$-wave pairing is a smooth crossover, for nonzero orbital-angular-momentum pairing such as $p$- or $d$-wave pairing, the BCS-BEC evolution is associated with a quantum phase transition at vanishing chemical potential $\mu$. In this paper, we have studied some generic features of the BCS-BEC quantum phase transition and the collective excitations in two-dimensional fermionic systems with $p$- and $d$-wave pairings. Our generic conclusions can be summarized as follows:
\\ (1) The quantum phase transition in two-dimensional fermionic systems is essentially related to the infrared behavior of the pairing interaction, which causes the infrared behavior of the fermionic excitation at $\mu=0$: $E_{\bf k}\sim k^l$, where $l=1$ for $p$-wave pairing and $l=2$ for $d$-wave pairing. The nonanalyticities of the thermodynamic quantities are due to the infrared divergence caused by the fermionic excitation at $\mu=0$.
\\ (2) The evolution of the Anderson-Bogoliubov mode is not smooth: Its velocity is nonanalytical across the quantum phase transition, due to the infrared divergence caused by the fermionic excitation at $\mu=0$.
\\ (3) The BCS-BEC quantum phase transition and nonsmooth evolution of the collective modes in 2D systems with nonzero orbital-angular-momentum pairing are solely related to the infrared behavior of the pairing interaction and are independent of the details of the interaction potential as well as the pairing symmetry.

Finally, we point out that while we have studied the 2D continuum models, the generic features of the BCS-BEC quantum phase transition summarized above also apply to fermions in a 2D square lattice \cite{lattice01,lattice02} since the infrared behavior of the fermionic quasiparticles remains. Therefore, it is interesting to extend our studies to lattice systems.

\emph{Acknowledgments} ---  G. Cao and P. Zhuang acknowledge the support from MOST(Grant No. 2013CB 92200). L. He is supported by the Helmholtz International Center for FAIR within the framework of the LOEWE program launched by the State of Hesse.

\end{document}